\documentclass[english,technote,onecolumn,12pt]{IEEEtran}
\usepackage{amstext}
\usepackage{newtxmath}
\usepackage[T1]{fontenc}
\usepackage[latin9]{inputenc}
\usepackage{color}
\usepackage{babel}
\usepackage{float}
\usepackage{units}
\usepackage{graphicx}
\usepackage[unicode=true,pdfusetitle,
 bookmarks=true,bookmarksnumbered=false,bookmarksopen=false,
 breaklinks=false,pdfborder={0 0 0},pdfborderstyle={},backref=false,colorlinks=true]
 {hyperref}
\hypersetup{
 linkcolor=blue,citecolor=blue,urlcolor=blue}

\makeatletter
\usepackage[font=normalsize]{caption}

\makeatother

\begin{document}
\title{Determination of the Optimum Square Resistance of Absorbing Vanes
of the Rotary Vane Attenuator{*}\thanks{\hspace*{-4mm}\textbf{\_\_\_\_\_\_\_\_\_\_\_\_\_}\medskip{}
\protect \\
{*}Invited paper}}
\author{Igor M. Braver, Pinkhos Sh. Fridberg, and Khona L. Garb}

\maketitle
\thispagestyle{empty}

The circular waveguide with an absorbing vane in the diametrical plane
serves as a rotor section of the rotary vane attenuator. The absorbing
vane is a thin dielectric substrate with a resistive film deposited
on it. Such a vane is supposed to provide the power attenuation to
the ${\rm TE}_{11}$ mode being polarized parallel to the resistive
film surface. The attenuation is due to the heat loss in the resistive
film according to the induced conduction current. The basic characteristic
parameter for the resistive film is its square resistance,
\begin{equation}
W=\frac{1}{\sigma d}\,,\label{eq:W}
\end{equation}
where $\sigma$ is the conductivity of the resistive material, while
$d$ is the resistive film thickness, which is small compared to the
skin depth.

It is evident that the greater the value of the linear attenuation
in the waveguide with a resistive film, the smaller the length of
the absorbing vane. Thus, to reduce the size of the rotary vane attenuator,
it is important to determine the value of $W$ when the attenuation
constant $q$ happens to be maximal. The existence of the peak of
the $q(W)$ function can be explained by the following physical considerations.
At $W\to0$, the film tends to be a perfect conductor ($\sigma\to\infty$)
that leads to zero power absorption. In the opposite limiting case
($W\to\infty$), the resistive film disappears ($d\to0$) and power
absorption is also zero. The aim of this paper is to determine the
optimum value of the square resistance $W$ that provides the maximum
linear attenuation in the waveguide with a resistive film.

\medskip{}

\textbf{\large{}Asymptotic Equations}{\large\par}

\medskip{}

Shimada \cite{Shimada1966} proposed an approximate method of calculation
of the attenuation constant for a circular waveguide with the resistive
film in the diametrical plane under the assumptions that $W/\zeta\gg1$,
where $\zeta=\unit[377]{\Omega}$ is the characteristic impedance
of free space, and that the resistive film does not distort the electromagnetic
field pattern of the ${\rm TE}_{11}$ mode, as shown in Figure \ref{fig:fields}a.
The expression derived by the author yields infinite values at the
cutoff frequency of the ${\rm TE}_{11}$ mode. This leads to large
errors when calculating $q$ at low frequencies of the waveguide operating
band. We derived a different equation, which is free from the above
mentioned shortcoming:
\begin{equation}
q=\frac{Q_{0}}{R\sqrt{2}}\sqrt{\sqrt{(k^{2}R^{2}-\varkappa_{11}^{2})^{2}+\left(\frac{TkR\zeta}{W}\right)^{2}}-k^{2}R^{2}+\varkappa_{11}^{2}},\qquad\frac{W}{\zeta}\gg1,\label{eq:q}
\end{equation}
where $Q_{0}=\unit[8.686]{dB}$, $\varkappa_{11}=1.841$, $T=1.026$,
$R$ is the waveguide radius, and $k=\frac{2\pi}{\lambda}$ with $\lambda$
being the wavelength in free space. In the frequency region differing
substantially from the cutoff frequency, equation (\ref{eq:q}) coincides
with Shimada's \cite{Shimada1966} equation.
\begin{center}
\begin{figure}[H]
\begin{centering}
\includegraphics{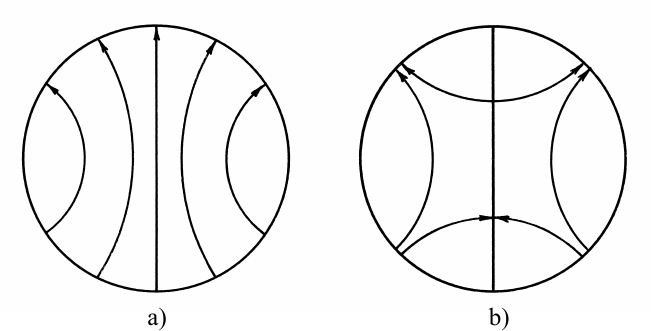}
\par\end{centering}
\caption{\label{fig:fields}Transverse electric field pattern in a waveguide
with a resistive film of large or small square resistance; (a) $W\to\infty$
and (b) $W\to0$.}

\end{figure}
\par\end{center}

We also derived an equation analogous to equation (\ref{eq:q}) for
a resistive film with a small value ($W/\zeta\ll1$) of the square
resistance. The distribution of the field of the mode being considered
in the waveguide with a low-resistance resistive film is shown in
Figure \ref{fig:fields}b, with the attenuation constant given by
the equation
\begin{equation}
q=\frac{Q_{0}}{R\sqrt{2}}\sqrt{\sqrt{(k^{2}R^{2}-\varkappa_{21}^{2})^{2}+\biggl(\left(MkR+\frac{N}{kR}\right)\frac{W}{\zeta}\biggr)^{2}}-k^{2}R^{2}+\varkappa_{21}^{2}},\qquad\frac{W}{\zeta}\ll1,\label{eq:q-2}
\end{equation}
where $\varkappa_{21}=3.054$, $M=1.175$, $N=19.793$.

\begin{figure}[H]
\begin{centering}
\includegraphics{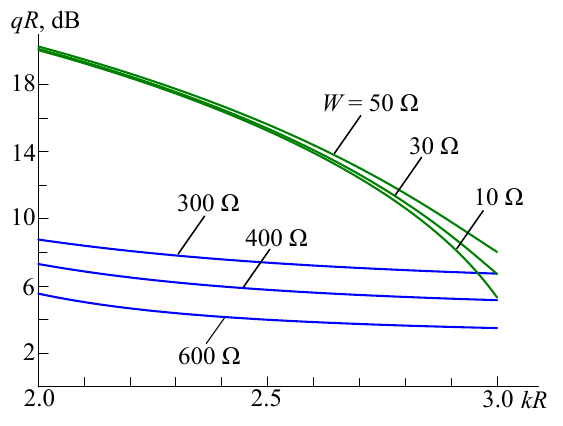}
\par\end{centering}
\caption{\label{fig:qR-kR}Attenuation constant vs. wavenumber. Curves for
$\unit[10]{\Omega}$ to $\unit[50]{\Omega}$ were calculated using
equation (\ref{eq:q-2}), while curves for $\unit[300]{\Omega}$ to
$\unit[600]{\Omega}$ were calculated using equation (\ref{eq:q}).}
\end{figure}

Representative plots of the asymptotic equations (\ref{eq:q}) and
(\ref{eq:q-2}) are shown in Figure \ref{fig:qR-kR}.
The attenuation constant as a function of frequency is decreasing.
Therefore, the absorbing vane must be designed to give the required
attenuation at the highest frequency of the waveguide operating band.
Qualitative aspects of the asymptotic equations show the existence
of a peak in the $q(W)$ dependence around the $W\sim\unit[100]{\Omega}$
region. For those square resistance values, the calculation procedure
should be performed using rigorous methods.

\medskip{}

\textbf{\large{}The Exact Solution}{\large\par}

\medskip{}

\begin{figure}[H]
\begin{centering}
\includegraphics{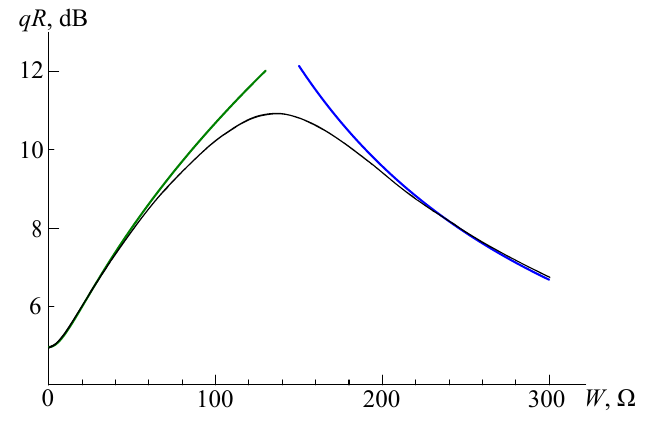}
\par\end{centering}
\caption{\label{fig:qr-W}Attenuation constant vs. square resistance of the
resistive film. The black curve shows the exact values, the blue curve
shows the approximate values calculated using (\ref{eq:q}), and green
curve shows the approximate values calculated using (\ref{eq:q-2}).}
\end{figure}

The general integral equation for the surface current for the problem
of the eigenmodes of a waveguide containing a resistive film has been
formulated in our work \cite{Braver1986}. In a subsequent paper \cite{Braver1986a},
this equation was used to investigate the dispersion properties of
a circular waveguide with a resistive film in the diametrical plane.
Taking into account the existence of the non-zero solution of the
integral equation, the dispersion equation to determine complex propagation
constants was derived. The numerical solution of this equation allows
for a highly accurate calculation of the linear attenuation of any
eigenmode of the circular waveguide with a resistive film. To design
the rotor section of a rotary vane attenuator, one needs to calculate
the attenuation constant of the mode possessing symmetry properties
of the ${\rm TE}_{11}$ mode polarized parallel to the resistive film
and propagating with the lowest degree of attenuation. As $W\to\infty$,
(the resistive film absent) this mode must transform into the ${\rm TE}_{11}$
mode. The results of the computation of the attenuation constant for
a number of values of $W$ are shown in Figure \ref{fig:qr-W}.
The computation was carried out at the highest frequency of the operating
band. Since the highest operating frequency is limited by the condition
that the ${\rm TE}_{21}$ mode be evanescent in the circular waveguide
($kR<\varkappa_{21}$), we take $kR=3$. The data calculated using
asymptotic equations (\ref{eq:q}) and (\ref{eq:q-2}) also are shown
in Figure \ref{fig:qr-W} for comparison. Thus, the plot in Figure
\ref{fig:qr-W} presents the solution of the problem under consideration,
that is, the peak linear attenuation is achieved in the case of the
square resistance equal to $130\div\unit[150]{\Omega}$.

\medskip{}

\textbf{\large{}The Matching Problem}{\large\par}

\medskip{}

\begin{figure}[H]
\begin{centering}
\includegraphics{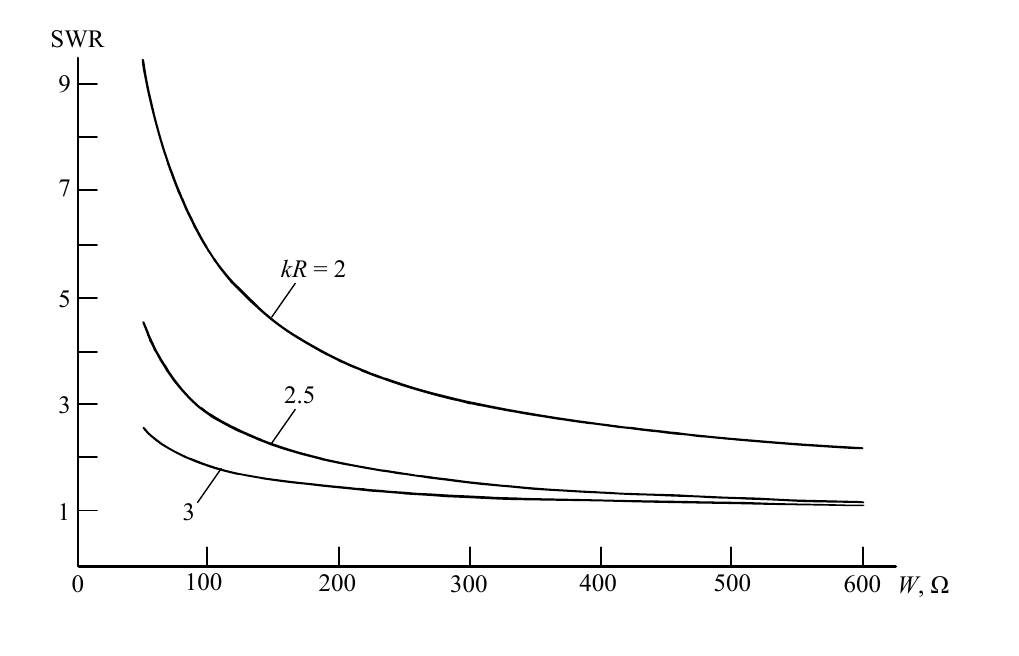}
\par\end{centering}
\caption{\label{fig:SWR}SWR vs square resistance of the resistive film.}
\end{figure}

The resistive film in the diametrical plane is a highly reflective
discontinuity for the ${\rm TE}_{11}$ mode being polarized parallel
to the resistive film surface. The corresponding diffraction problem
was solved in our works \cite{Braver1990,Braver1990a}, and the technique
for calculating the reflection coefficient $\Gamma$ was developed.
The computed plots representing the standing wave ration ${\rm SWR}=(1+\Gamma)/(1-\Gamma)$
as a function of the resistive film square resistance are shown in
Figure \ref{fig:SWR}. The computation was performed with the resistive
film length --- the size along the waveguide axis --- set to $2R$.
Further increase of the resistive film length does not actually lead
to the change of the SWR values as the resistive film tends to cause
high attenuation ($\sim\unit[30]{dB}$).

The resistive film with the optimum square resistance of $130\div\unit[150]{\Omega}$
requires the additional impedance matching. As the resistive film
with $W\sim\unit[400]{\Omega}$ is characterized by substantially
lower SWR values, the following techniques for producing an absorbing
vane are recommended. The first side of the dielectric substrate should
to be deposited with the resistive film of square resistance $W_{1}=\unit[400]{\Omega}$,
having short tapered sections at both ends. The second side of the
substrate should to be deposited with the resistive film of square
resistance $W_{2}=\unit[200]{\Omega}$ at a short distance from the
ends of the vane. The resistive film on this side should also have
short tapered endings.
The effective square resistance of the middle part of the vane has
the required value of $\approx\unit[130]{\Omega}$, calculated by
\begin{equation}
W_{{\rm ef}}=\frac{W_{1}W_{2}}{W_{1}+W_{2}}.
\end{equation}

\begin{figure}[H]
\begin{centering}
\includegraphics{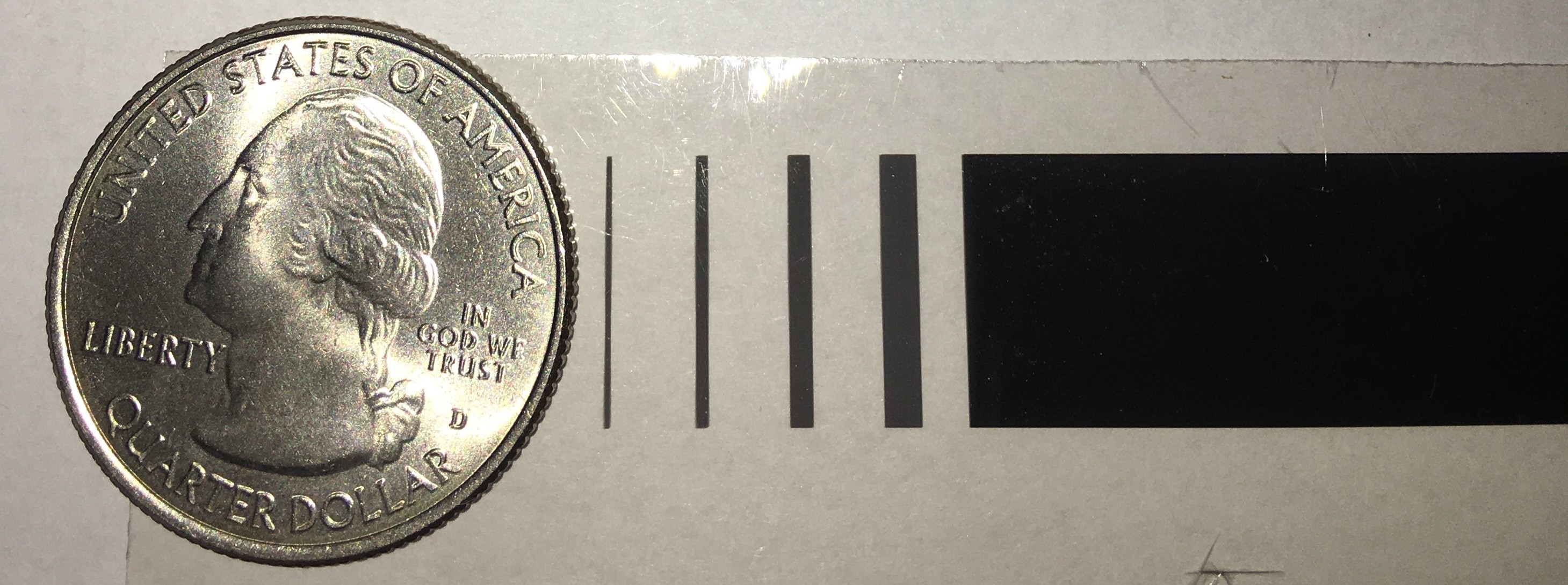}
\par\end{centering}
\caption{\label{fig:coin}Low-reflectivity absorbing vane.}
\end{figure}

Another type of impedance matching technique uses narrow resistive
strips leading the resistive film (see Fig.~\ref{fig:coin}) with
the optimum value of the square resistance. The length of each strip
and the distance between strips are calculated by the numerical optimization
methods on the basis of a computer program used to determine the scattering
matrix for the resistive film of arbitrary length. Computations showed
that in order to decrease the SWR value from 5 to 1.05 over the whole
frequency band of the waveguide operating frequencies, it is sufficient
to deposit four narrow resistive strips leading the resistive film
with $W=\unit[130]{\Omega}$. Those additional strips should have
the square resistance value $W\sim200\div\unit[400]{\Omega}$. In
practice, it may be more convenient for the additional strips and
the main one to have the same resistance $W=\unit[130]{\Omega}$.
In that case, five leading strips should be deposited. Experimental
samples of absorbing vanes with low reflectivity (Fig.~\ref{fig:coin})
were produced according to our calculations at the Vilnius Scientific
Research Institute of Radio Measuring Instruments, Lithuania.

\section*{Conclusion}

The attenuation constant in the circular waveguide with a resistive
film in the diametrical plane decreases with frequency. The peak of
the attenuation constant as the function of the square resistance
at the highest frequency of the operating band is determined to be
at $W=130\div\unit[150]{\Omega}$. The use of the resistive film with
the square resistance mentioned above provides the required attenuation
at the shortest possible length of the absorbing vane.

\end{document}